\documentclass[pra,aps,twocolumn,groupedaddress,showpacs,floatfix,preprintnumbers]{revtex4}
\usepackage{amsmath,amssymb,multirow,epsfig,bm,color}

\newcommand{\beq}{\begin{equation}}
\newcommand{\eeq}{\end{equation}}
\newcommand{\bea}{\begin{eqnarray}}
\newcommand{\eea}{\end{eqnarray}}

\newcommand{\eF}{\varepsilon_{F}}

\newcommand{\kF}{k_{\textrm{F}}}

\newcommand{\ferron}{{\it ferron}}
\newcommand{\alphup}{\alpha_{\uparrow}}
\newcommand{\alphdw}{\alpha_{\downarrow}}
\newcommand{\nup}{n_{\uparrow}}
\newcommand{\ndw}{n_{\downarrow}}
\newcommand{\taup}{\tau_{\uparrow}}
\newcommand{\tadw}{\tau_{\downarrow}}

\begin{document}

\title{
Spin-polarized droplets in the unitary Fermi gas
}

\author{Piotr Magierski$^{1,2}$}\email{piotrm@uw.edu}
\author{Bu\u{g}ra T\"uzemen$^{1}$}\email{tuzemen@if.pw.edu.pl}
\author{Gabriel Wlaz\l{}owski$^{1,2}$}\email{gabriel.wlazlowski@pw.edu.pl}

\affiliation{$^1$Faculty of Physics, Warsaw University of Technology, Ulica Koszykowa 75, 00-662 Warsaw, Poland}
\affiliation{$^2$Department of Physics, University of Washington, Seattle, Washington 98195--1560, USA}

\begin{abstract} 
We demonstrate the existence of a new type of spatially localized excitations 
in the unitary Fermi gas: spin polarized droplets with a peculiar internal structure involving 
the abrupt change of the pairing phase at the surface of the droplet. 
It resembles the structure of the Josephson-$\pi$ junction occurring
when a slice of a ferromagnet is sandwiched between two superconductors. 
The stability of the impurity is enhanced by the mutual interplay between
the polarization effects and the pairing field, resulting in an exceptionally 
long-lived state. The prospects for its realization in experiment are discussed.
\end{abstract}


\maketitle

\section{\label{sec1}Introduction}

The unitary Fermi gas, routinely realized in ultracold atomic gases, 
is a remarkable system possessing universal properties. One of its features is an exceptionally strong pairing which results in a pairing gap of the order of half of the Fermi energy~\cite{bloch}. 
Such a strong pairing field makes it an appealing system for studies of pairing-related phenomena. 
In particular, the spin-imbalanced systems offer the possibility to investigate the superfluidity in 
the time-reversal symmetry breaking conditions.
Due to the different radii of Fermi spheres for spin-up and spin-down fermions one expects the appearance 
of effects characteristic for the FFLO phase~\cite{ff,lo}, Sarma phase~\cite{sarma,gubbels,pao,hu} or interior gap phase~\cite{wilczek}.
Unfortunately in the trapped inhomogeneous gases the excess of the majority-spin
particles are expelled towards the edges of the cloud, and the predicted effects of nonstandard pairing mechanisms 
appear close to the trap boundaries~\cite{castorina2005,kinnunen2006,machida2006}, which make them difficult to observe experimentally.
However, the recent experimental realization of the box-like traps offers the possibility to investigate almost homogeneous systems,
and thus may eventually help in detection of these exotic phases~\cite{box-trap}.

In spin-imbalanced systems the behavior of the pairing field at the interface between superfluid (paired atoms) and normal (excess of spin majority atoms) component is similar to the one at the superconductor(S)-ferromagnet(F) junction~\cite{buzdin}.
The issue which however is of paramount importance in the case of ultracold atoms, and makes it qualitatively different from 
the solid state physics, is the stability and dynamics of these structures.
Namely, in the ultracold atomic gas the stability is governed by the same interaction which is 
responsible for its pairing properties. This is quite a different situation than in the case of the solid state structures, where the long-range Coulomb interaction between heavy ions governs the stability of e.g. SF junction and the pairing properties 
of electrons contribute to transport and thermal properties only.
In this letter we describe a novel type of long-lived structure in ultracold atomic gas which appears when the spin-polarization is induced locally by an external potential. 

\section{\label{sec2}Spin polarized droplets}

\begin{figure}[b]
   \begin{center}
   \includegraphics[width=\columnwidth, trim=0 0 0 00, clip]{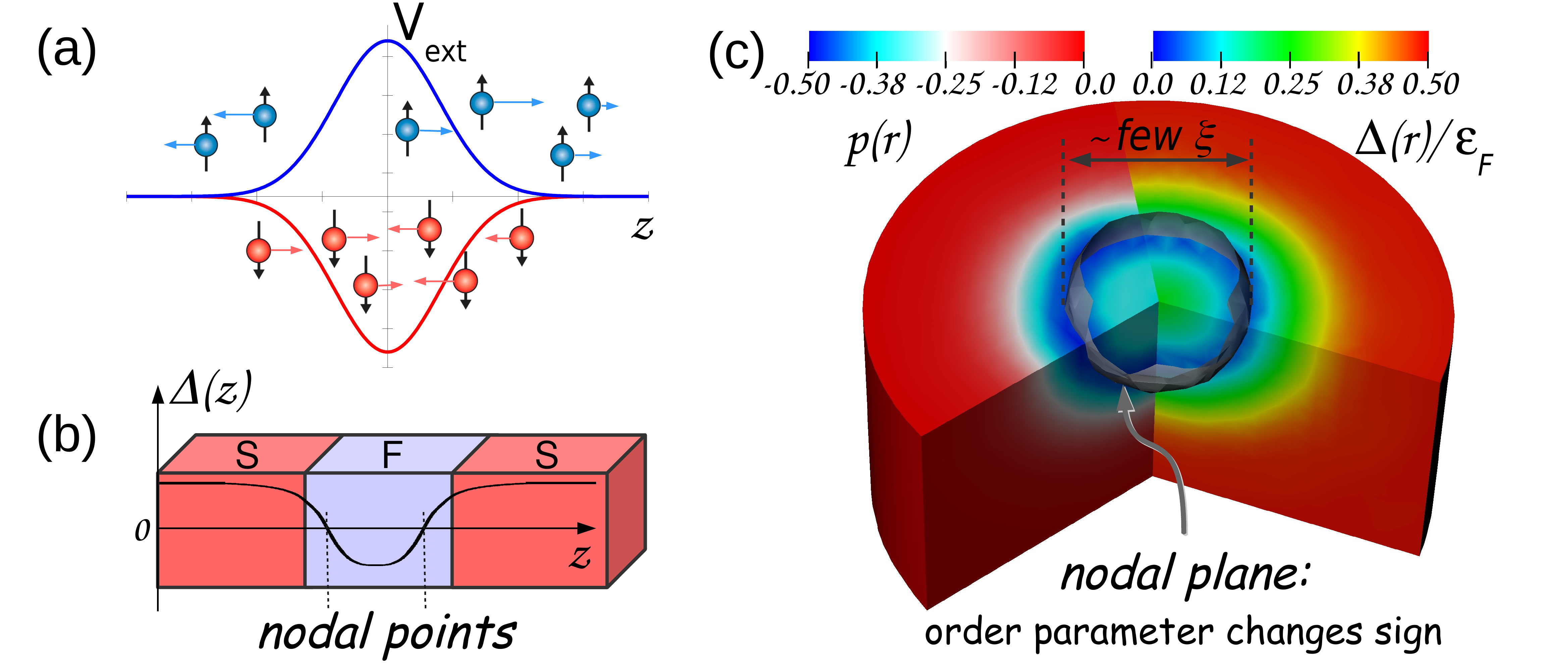}
   \end{center}\vspace{-3mm}
   \caption{(a)~The spin-polarizing potential consisting of two Gaussian-shape potentials that couple with different signs to different spin states. The potential generates the region where the system is locally polarized $n_{\uparrow}(\bm{r})\ll n_{\downarrow}(\bm{r})$.
   (b)~Schematic structure of the order parameter $\Delta$ inside SFS junction, where the ferromagnet layer is sufficiently thin. Inside the layer the order parameter is suppressed and changes sign. 
   (c)~Structure of the polarized droplet. 
The left and right parts of the subfigure show the spin polarization $p(\bm{r})$ and the pairing field $\Delta(\bm{r})$ distributions, respectively. The characteristic feature is the presence of the nodal surface of the pairing field 
at which pairing changes its sign. The spin polarization reaches maximum in the vicinity of the nodal surface.}
   \label{fig:1}
\end{figure}

\begin{figure}[b]
   \begin{center}
   \includegraphics[width=\columnwidth, trim=80 40 80 20, clip]{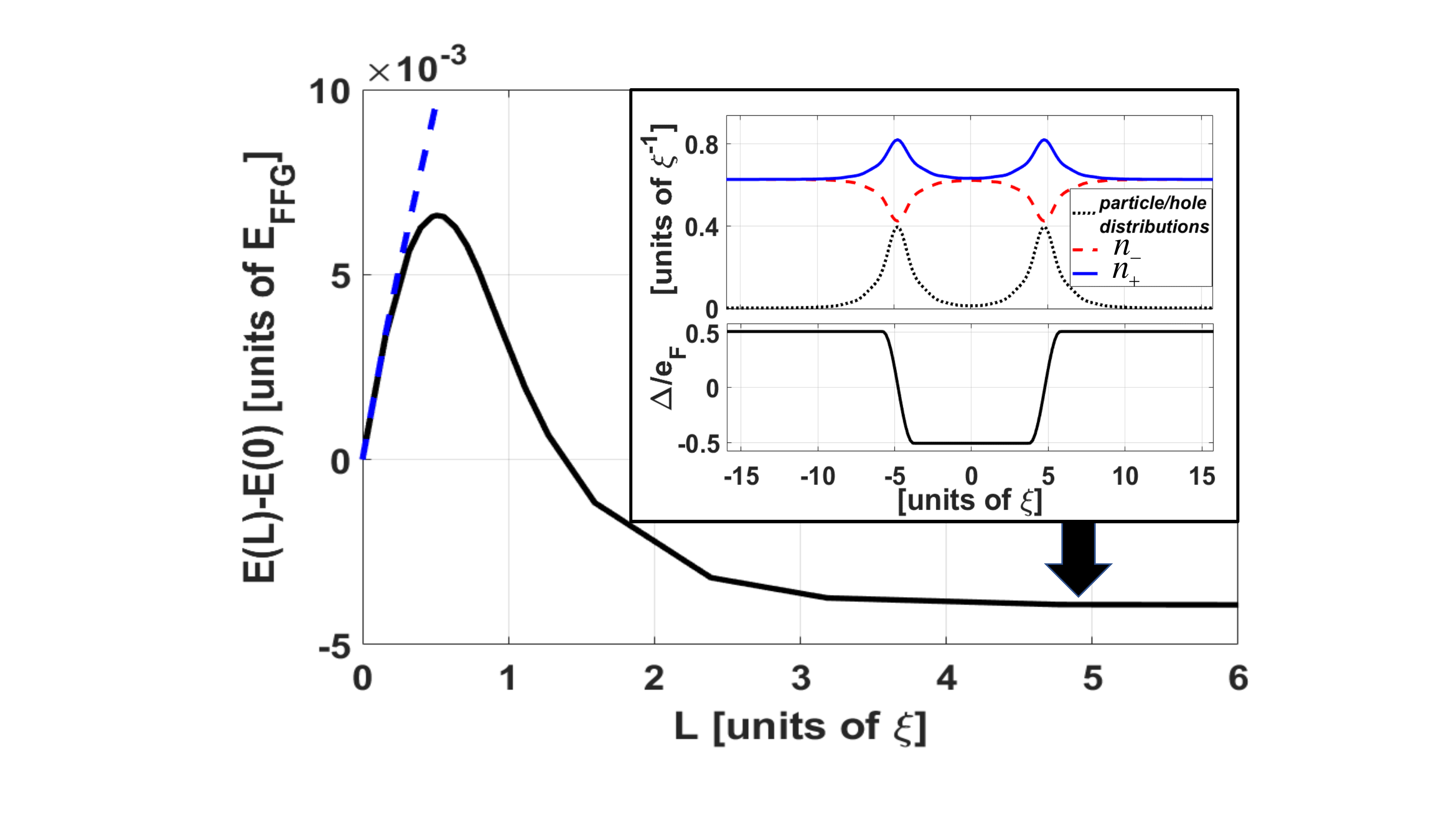}
   \end{center}\vspace{-1mm}
   \caption{
   The total energy of 1D Fermi gas with (total) polarization $p=0.05$ in an external pairing field,
   as a function of the distance between pairing nodal points ($2L$). The dashed line
   indicates the energy of the configuration with pairing set to zero between nodal points.
   The inset shows the pairing configuration (lower subfigure) at the distance indicated by
   the arrow. The upper subfigure shows the density distributions of spin-up and spin-down fermions
   and the particle(spin-up)/hole(spin-down) distributions associated with Andreev states localized 
   around the nodal points, which are responsible for spin-polarization. 
   $E_{FFG}$ denotes the energy of the uniform, unpolarized Fermi gas with the same number 
   of particles, $\xi$ is the coherence length of the corresponding uniform system. 
}
   \label{fig:2}
\end{figure}

Let us consider the unpolarized and uniform unitary Fermi gas (UFG). When the time-dependent and spin-polarizing potential is applied 
within a certain region and it is strong enough, it locally breaks Cooper pairs, see Fig.~\ref{fig:1}(a).
Namely, it creates a region where the pairing field is weaker (or even vanishes at all) and is characterized by a nonzero spin-imbalance 
$p(\bm{r})=\frac{n_{\uparrow}(\bm{r})-n_{\downarrow}(\bm{r})}{n_{\uparrow}(\bm{r})+n_{\downarrow}(\bm{r})}$, where $n_{\uparrow(\downarrow)}$ is the density of spin-up (down) fermions. 
Consequently due to the different locations of Fermi surfaces of spin-up and spin-down fermions, 
induced by the polarization, the pairing field starts to oscillate and changes sign inside the impurity. 
Similar situation is encountered in SFS junctions~\cite{buzdin, Halterman}. 
In Fig.~\ref{fig:1}(b) we present a sketch of behaviour of the order parameter $\Delta$ through the SFS junction within a thin ferromagnet layer. 
Regions close to the nodal points, where $\Delta\rightarrow 0$, store the unpaired particles~\cite{SuppressedSolitonicCascade}, and in these regions the spin-polarization is enhanced.
This effect may be viewed also as a consequence of occupation of certain Andreev states which
are localized around the pairing nodal points due to the scattering of quasiparticles
on a spatially varying pairing potential (see also Fig.~\ref{fig:2}). As a consequence the pairing nodal points and the enhanced spin-polarized regions are mutually connected.
While aforementioned effects are known it is surprising that the structure persists even if one removes the external potential. 
This self-sustained polarized droplet is presented in Fig.~\ref{fig:1}(c).

The natural question concerns origin of the stability of the structure, which as we will argue in this Letter is particularly enhanced.
If we ignore the pairing structure inside the impurity its stability is governed by the
spin transfer processes which result from the scattering of quasiparticles on the interface
between the superfluid and spin-polarized region. If the system is at low temperatures and
not far from equilibrium the spin transfer, which in this regime is mainly due to Andreev reflection, 
is effectively suppressed~\cite{bert2007, parish2009, sommer2011, enss2012}.
However the creation of the polarized impurity by time dependent potential brings the system
relatively far from equilibrium. The process of inducing polarization creates various phonon excitations
in a superfluid, which would eventually decay into quasiparticle excitation and increase the temperature 
of the system. This effect would speed-up the spin transfer processes between polarized region
and the superfluid surrounding, leading eventually to the disappearance of the impurity.
When pairing is present inside the impurity the proximity effects, similar to those in SFS junction, 
enhance stability of the configuration.
In order to understand better the reason of the stability it is instructive
to consider the 1D case. In this case the initially created structure starts to expand as 
the two polarized regions, in the vicinity of nodal points, repel each other.
This in turn implies that the pairing field becomes complex: $\Delta \rightarrow \Delta\exp(i q x)$ and
induces the current $|\vec{j}| \propto q$. As a result one ends up with two polarized regions which travels with finite velocity in opposite directions and the region between them becomes again fully paired, see Fig.~\ref{fig:2}. For more details of 1D calculations see Appendix~\ref{appH}.
In the 3D case this process however will not occur, as the radial expansion of the polarized region, 
would inevitably increase the size of the polarized shell which clearly cost energy.
Thus the ``surface tension'' associated with the polarized shell counterbalances the repulsion and stabilizes size of impurity.
Namely, the stability of the polarized impurity is dictated by the energy balance of the impurity energy, which can roughly be
decomposed in two terms:
\begin{equation}
E_{imp} = E_{shell} + E_{int},
\end{equation}
where $E_{shell}$ describes the energy of the polarized shell surrounding the impurity (concentrated 
around the pairing nodal surface) and $E_{int}$ is the energy related to the impurity volume.
The interplay between these terms lead to an enhanced stability in 3D.
Similarly, the opposite process leading to the collapse of the impurity is suppressed due to
the presence of the pairing field inside with an opposite phase. It creates an effective potential barrier as the 
pairing inside has to be destroyed during the collapse.
We emphasize that this configuration correspond to an excited state. We dub it as \ferron, since the pairing field inside the impurity and in SFS junction share many similarities.

The considerations concerning the stability of the polarized region allows for the estimations
of the minimum impurity size.
It is clear that the pairing inside the impurity needs a space to develop which is of the order
of the BCS coherence length $\xi$. This sets the minimal limit for the size of the impurity to be stable.
The numerical simulations admit that minimal radius of the stable ferron is about $2.5\,\xi$. 


\section{\label{sec3}Numerical simulations}

\subsection{\label{sec3A}Framework}

In order to investigate the effect described above the unconstrained 3D numerical calculations
has been performed, using the framework based on time dependent density functional theory in the form of asymmetric superfluid local density approximation (TDASLDA). It allows for an accurate treatment of the pairing correlations in time-dependent fashion. It employs the density functional of generic form (we set units to $m=\hbar=k_B=1)$\cite{LNP__2012}:
\begin{eqnarray}\label{func}
\mathcal{E}_{\small{ASLDA}} &=& \alphup(p)\frac{\taup}{2} +  \alphdw(p)\frac{\tadw}{2} + \beta(p)(\nup+\ndw)^{5/3}\nonumber \\
& & + \frac{\gamma(p)}{(\nup+\ndw)^{1/3}}\nu^{\dagger}\nu + \sum_{i=\uparrow,\downarrow}[1-\alpha_{i}(p)]\frac{\bm{j}^2_{i}}{2n_{i}}.
\end{eqnarray}
The terms $n_{i}$, $\tau_{i}$, $\nu$ and $\textbf{j}_{i}$ respectively denote the normal, kinetic, anomalous and particle current densities which are defined in terms of Bogoliubov quasi-particle wavefunctions$\{ v_{n,i}, u_{n,i} \}$:
\begin{eqnarray}
n_{i}(\textbf{r}) &=& \sum_{|E_n|<E_c} |v_{n,i}(\textbf{r})|^2 f_{\beta}(-E_n), \\
\tau_{i}(\textbf{r}) &=& \sum_{|E_n|<E_c} |\nabla v_{n,i}(\textbf{r})|^2 f_{\beta}(-E_n), \\
\nu(\textbf{r}) &=& \sum_{|E_n|<E_c}  v_{n,\downarrow}^{*}(\textbf{r}) u_{n,\uparrow}(\textbf{r})\frac{f_{\beta}(-E_n)-f_{\beta}(E_n)}{2}, \\
\textbf{j}_{i}(\textbf{r}) &=& \sum_{|E_n|<E_c} \textrm{Im}[v_{n,i}(\textbf{r}) \nabla v_{n,i}^{*}(\textbf{r})]f_{\beta}(-E_n),
\end{eqnarray}
where $E_n$ is the quasiparticle energy and $E_c$ is the energy cut-off value as required by the regularization scheme. 
The Fermi-Dirac distribution, $f_{\beta}(E) = 1/(\exp(\beta E)+1)$, where $\beta=1/T$, allows to model finite temperature effects. The terms in the functional given by Eq. (\ref{func}) have the following meaning: 
the first two describe kinetic energies of particles with spin $i=\{\uparrow,\downarrow\}$ possessing effective masses $\alpha_{i}$, 
the third and the fourth terms describe 
normal and pairing interactions, respectively where strengths are controlled by coefficients $\beta$ and $\gamma$. The last two terms are required in order to preserve Galilean invariance of the theory. All coupling constants $\alpha_{i}$, $\beta$ and $\gamma$ are functions of local polarization of the gas $p=\frac{\nup-\ndw}{\nup+\ndw}$. They have been adjusted to quantum Monte Carlo results 
for spin-imbalanced, homogeneous unitary Fermi gas and exhibits a remarkable agreement with the calculations for trapped systems~\cite{LNP__2012,ASLDA-LOFF}. In this way 
the density functional treatment offers a description of superfluidity beyond the meanfield Bogolubov-de Gennes (BdG) approximation which is unable to reproduce correctly the quantum Monte Carlo data.

The TDASLDA equations can be obtained from the stationarity condition of the action:
\begin{equation}
S=\int_{t_0}^{t_1} \left( \langle 0(t)| i\frac{d}{dt}| 0(t) \rangle - E(t) \right) dt,
\end{equation} 
where $|0(t) \rangle$ denotes the quasiparticle vacuum at time $t$ and $E(t)$ is the total energy 
\begin{equation}
E(t)=\int \left ( \mathcal{E}_{\small{ASLDA}}(\bm{r},t) + \sum_{i=\uparrow,\downarrow} V_{i}(\bm{r},t)n_{i}(\bm{r},t) \right ) d\bm{r}.
\end{equation} 
$V_{i}$ is an arbitrary
external one-body potential, which couples to the number density $n_{i}$. 
Formally TDASLDA equations resemble the time dependent Bogoliubov de-Gennes (TDBdG) equations.
\begin{equation}
i\frac{\partial}{\partial t} \left( \begin{array}{cc}
u_{n,\uparrow}(\textbf{r},t) \\
v_{n,\downarrow}(\textbf{r},t)
\end{array} \right) = \left( \begin{array}{cc}
 h_{\uparrow}(\textbf{r},t) & \Delta(\textbf{r},t) \\
 \Delta^*(\textbf{r},t) & -h_{\downarrow}^*(\textbf{r},t)
\end{array} \right)\left( \begin{array}{cc}
u_{n,\uparrow}(\textbf{r},t) \\
v_{n,\downarrow}(\textbf{r},t)
\end{array} \right).
\end{equation}
Here $h_{i}(\textbf{r},t)$ denotes the single particle hamiltonian, which consists of the kinetic, mean-field and external potential terms. The pairing field $\Delta$ is proportional to the anomalous density $\nu$. Spin reversed components of quasi-particle wavefunctions can be obtained via symmetry relation $u_{n,\uparrow}\rightarrow v_{n,\uparrow}^*$,  $v_{n,\downarrow}\rightarrow u_{n,\downarrow}^*$ and $E_n \rightarrow -E_n$.

The framework has been extensively tested over last years, and it proved 
to provide an accurate description of various dynamical properties of the strongly interacting Fermi gas, including generation and proliferation of quantum vortices~\cite{Science__2011} or dynamics of solitonic cascades~\cite{SuppressedSolitonicCascade,PRL__2014}. More details related to technical aspects of solving TDASLDA equations we provide in the Appendix~\ref{appA}.
 
\subsection{\label{sec3C}Demonstration of droplet-like properties of the ferron}
The initial condition for simulations consists of the uniform solution of unpolarized
unitary Fermi gas, at very low temperature $T/\eF=0.01$, where $\varepsilon_F$ denotes the Fermi energy. 
In order to check the stability of the results with respect to finite size effects, we used the lattice sizes 
ranging from $40^3$ to $64^3$. The lattice constant $dx$ is chosen to be about three times smaller than the average interparticle 
distance $n^{-1/3}$, and the number of particles is fixed to get $k_{F} dx\approx 1$, where $\kF=(6\pi^2 n_{\uparrow})^{2/3}$ is Fermi wave vector. Subsequently the external potential in the form of the gaussian $V_{i}(\textbf{r},t) = \lambda_{i} A(t) e^{-r^2/2\sigma^2}$
of width $\sigma$ and the amplitude $A_{0}=\max[A(t)]$
has been applied. The potential is spin-dependent, attracting spin-down atoms 
($\lambda_{\downarrow}=-1$) and repelling spin-up atoms ($\lambda_{\uparrow}=1$), as shown in Fig.~\ref{fig:1}(a).
The potential is switched on for a time interval, sufficient to induce locally spin polarization, 
leading to the destruction of Cooper pairs within the region of radius $R\approx \sigma$.
Eventually the external potential is removed, and the system evolves in time. For more details of protocol for the impurity generation see Appendix~\ref{appB}.
\begin{figure}[b]
   \begin{center}
   \includegraphics[width=\columnwidth, trim=0 150 0 220, clip]{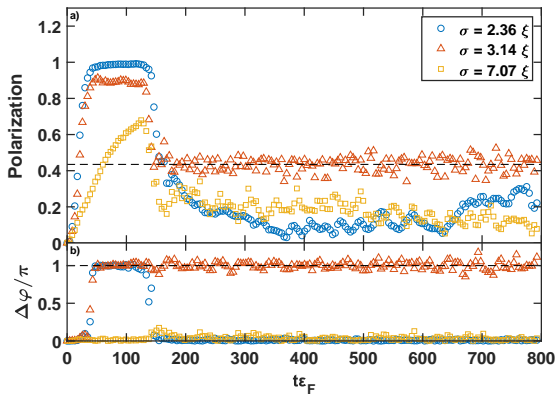}
   \end{center}\vspace{-4mm}
   \caption{Time evolution of the local polarization $p(\bm{r}=0)$ in center of the box (panel a), and the pairing phase difference $\Delta\varphi$ (panel b). The external potential of strength $A_0=2\eF$ and various widths $\sigma$ has been applied within time interval $0-150\eF^{-1}$. In the case $\sigma=3.14\xi$ the
stabilization of the polarization and the phase difference at constant value for $t\eF>150$ indicates 
creation of the ferron. For animations showing distributions of both local polarization $p(\bm{r})$ and 
pairing field $\Delta(\bm{r})$ see movies attached to Supplemental Material~\cite{Supplemental}}
   \label{fig:MA_polar_phase}
\end{figure}
In Fig.~\ref{fig:MA_polar_phase} the time evolution of the local polarization, induced
at the center of the simulation box $p(\bm{r}=0)$, is presented. In the lower panel
the relative phase of the pairing field at the center of the polarized region, measured with 
respect to the phase outside the impurity  $\Delta\varphi = \varphi_{\textrm{in}}-\varphi_{\textrm{out}}$ is shown, where $\varphi = \arg \Delta$.
Simulations were performed for the external potential of the amplitude $A_0=2\eF=\kF^2$ 
and three widths $\sigma/\xi = 2.38, 3.14, 7.07$.
The potential has been applied within time interval $t<150\eF^{-1}$. In 
the first two cases the time interval was sufficiently long to generate the region 
of high local polarization $p\gtrsim 0.8$ with almost vanishing paring field $\Delta(\bm{r})\approx 0$ inside the region. The evolution after the potential was removed turned out to be different
and depended crucially on the size of the polarized region. Too small polarization radius
$\sigma\lesssim 3\xi$ does not provide enough space to develop aforementioned
oscillatory pairing field pattern. As a result it is not stabilized by the pairing field and decays. 
For larger sizes, $\sigma\gtrsim 3\xi$, as the potential 
is removed, the pairing field inside the impurity reappear with the opposite phase. Moreover, within
the time scale $1,000\eF^{-1}$ there is no visible decrease
of the polarization indicating the unusual stability of the polarized droplet. In the last case, for $\sigma\approx 7\xi$, 
the potential acts within too short time interval to excite the ferron. 
Thus, both the potential radius and its duration have to be large enough
to generate the stable polarized droplet. It turns out that in the last scenario 
the time interval needs to be increased at least by a factor of $2$ in order to produce 
the self-sustained spin-polarized droplet. 

The numerical simulations indicate that there is a preferable shape and size of the impurity. 
The calculations reveal that changing 
the potential width $\sigma$ within the range $3\xi-7\xi$ weakly affects 
the size of the generated ferron. In each case the measured radius of the droplet 
(defined as a distance to the nodal surface) evolves 
towards a value within the interval $2.5\xi-3.5\xi$. 
Another droplet-like feature of the ferron can be observed when a deformed impurity is generated.
Namely, the external potential in the form of the gaussian function $V_{i} = \lambda_{i} A\exp\left(-\frac{x^2}{2\sigma_x^2}-\frac{y^2}{2\sigma_y^2}-\frac{z^2}{2\sigma_z^2}\right)$ with different widths $\sigma_{x,y,z}$ 
in each spatial direction induces initially a deformed impurity.
However it evolves rather fast (within time of order $~100\eF^{-1}$) towards almost spherically symmetric configuration. 
These results confirm that the peculiar pairing structure is responsible for
the existence and stability of the ferron. Spherical shape of the droplet at fixed 
volume minimizes $E_{shell}$, without affecting $E_{int}$, whereas the preferable size 
of the impurity is due to the nodal structure of the pairing field. 

\begin{figure}[h]
   \begin{center}
   \includegraphics[width=\columnwidth, trim=0 0 0 0, clip]{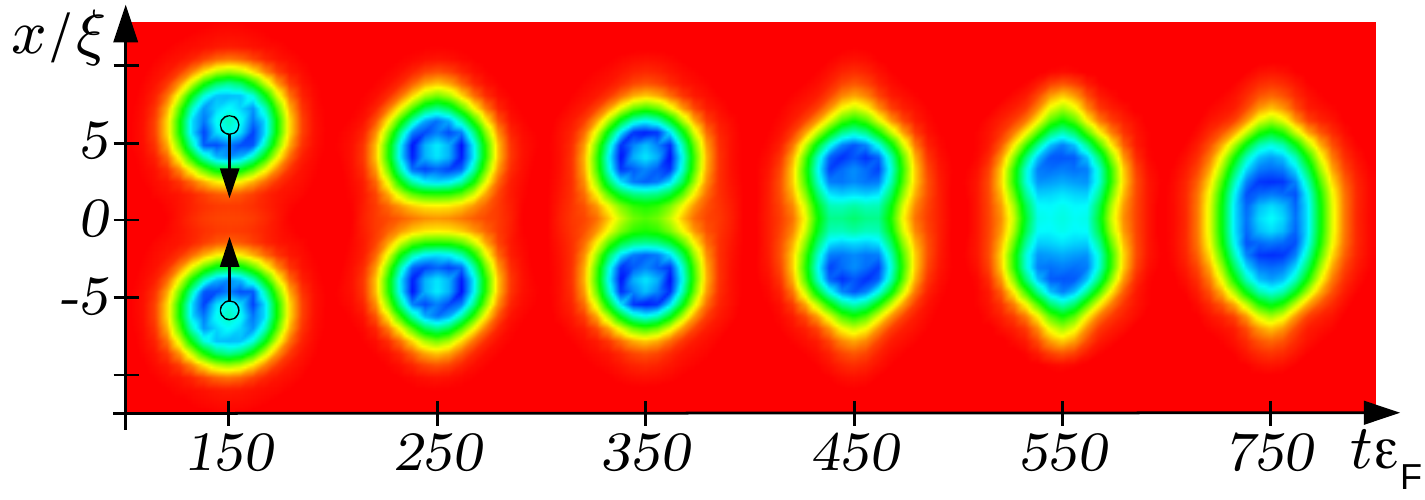}
   \end{center}\vspace{-3mm}
   \caption{Head-on collision of two ferrons. The droplets were created by two moving potentials of amplitude $A = 2\eF$ and width $\sigma = 3.14\xi$ as indicated by arrows. The potentials were applied within time interval $0-150\eF^{-1}$, and snapshots of pairing field spatial distributions $\Delta(\bm{r})/\eF$ are taken at 
later times (as indicated by labels on horizontal axis). Color coding is the same as in Fig.~\ref{fig:1}.
For full movie see Supplemental Material~\cite{Supplemental}.}
   \label{fig:4}
\end{figure}

The above simulations demonstrate that ferrons~in 3D form long-lived excitations
and indeed bear similarities to droplets. 
In order to investigate further this aspect the collisions of two initially 
separated ferrons have been performed. 
They have been generated by applying two spin-selective potentials moving towards each other. 
The potentials' velocities, generating ferrons, were set smaller than 
the speed of sound and the process of head-on collision is shown in Fig.~\ref{fig:4}. 
Note that the structure of ferrons is preserved
during the collisions. They fuse
and a new droplet of bigger size is created with the typical pairing nodal structure. 
Although initially the obtained ferron is deformed one expects
it will evolve toward the spherical shape, which may take a relatively long time, as it is immersed
in an excited superfluid bath.

Finally, we have also checked that qualitatively the same results are obtained within mean-field BdG approach, see the Appendix~\ref{appI} for comparison. 
It means the existence of the stable ferron is not sensitive to choice of a particular form of the functional. 

\section{\label{sec44}Stability of the results with respect to perturbations}

\subsection{The impurity stability vs its size} \label{appC}
\begin{figure}[t]
\includegraphics[width=\columnwidth, trim=0 160 0 160 clip]{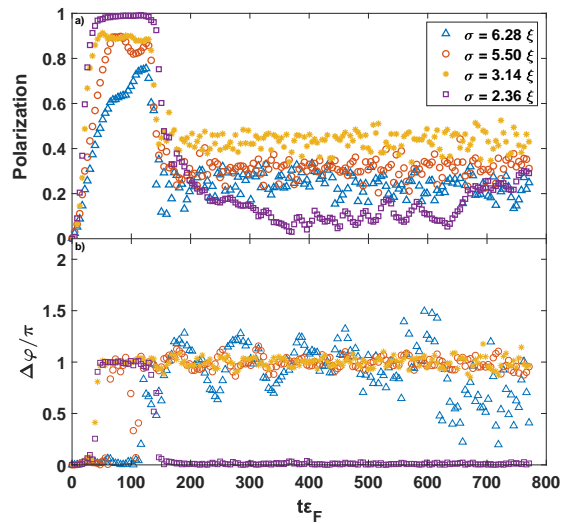}
\caption{Impact of the spin polarizing potential width $\sigma$ on the process of the impurity generation. Panel a) shows time evolution of the local polarization in the center of the impurity, while panel b) shows the pairing phase difference. The box size: $40^3$, corresponds to the length of $31\xi$ along 
each dimension. The amplitude of the potential was fixed at $A_0=2\eF$. Results for $\sigma\gtrsim 3\xi$ corresponds to the cases where the ferron has been successfully created. 
Calculations presented on this graph are visualized in {\it Movies 1-4}~\cite{Supplemental}.\label{SMfig:pol_phase}}
\end{figure}

The width $\sigma$ of the external spin polarizing potential, Eq.~(\ref{SMeq:V_p}), represents an important parameter 
in the process of the droplet generation. If it is too small, then inside the polarized region there is not enough space 
to allow for the order parameter fluctuations, which naturally occur at the scale of the coherence length ($\xi$).
Therefore, based on the presented argument, one expects that $\sigma$ has to be at least of the order of $\xi$.
Numerically we have confirmed that the minimal width needed for a successful creation of the ferron is indeed
$\sigma\gtrsim 3\xi$, which is a threshold value (as we go through the impurity the phase difference varies from 
$0$ to $\pi$ and again to $0$). 
In Fig.~\ref{SMfig:pol_phase} we presented the time evolution of the spin polarization inside the impurity and the phase difference between 
the interior and the outside region, for different values of $\sigma$. It is clearly seen that for widths exceeding the coherence length $\xi$ by factor $3$, 
both the polarization and the phase difference remains fairly constant during the simulations, which correspond to times 
$t\eF\approx 1000$. 
Note that as we increase the width of the potential we excite more phonons in the background superfluid. 
These fluctuations propagate towards edges of the box and due to periodic boundary conditions they again re-enter the box and interfere 
with the ferron structure. This effect, which is due to finite size of the box and imposed periodic boundary conditions,
gives rise to oscillations of the phase difference, especially well visible for case with $\sigma=6.28\xi$.  
\begin{figure}[t]
\includegraphics[width=\columnwidth, trim=0 140 0 140, clip]{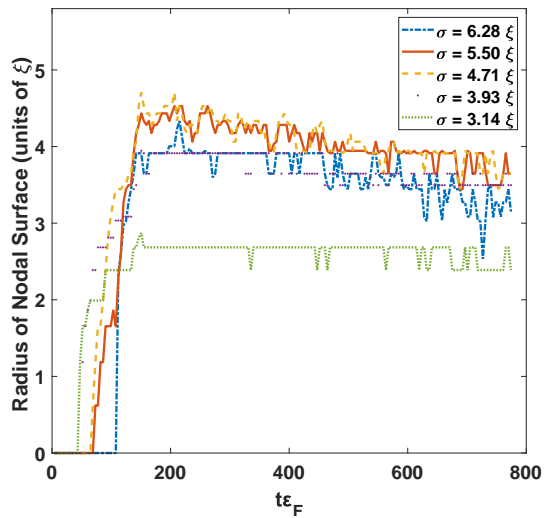}
\caption{Time evolution of radii of impurities generated by potentials of different widths $\sigma$. The amplitude of the potential was fixed at $A_0=2\eF$. 
The potential was turned off at time $t_{\textrm{off}}=150\eF^{-1}$. 
Variations of radii at short time scales are related to the finite resolution of the 
lattice.\label{SMfig:radius}}
\end{figure}

Performed calculations indicate the existence of a preferable size of the impurity. In order to demonstrate this feature, 
let us consider certain volume $V_{\pi}$ characterized by the shifted phase (by $\pi$) of the order parameter with respect to the value 
far away from the polarized region. This volume may serve as defining the radius of the ferron by means of the relation $V_\pi=\frac{4}{3}\pi r^3$. 
In Fig.~\ref{SMfig:radius} we presented the radii of impurities, generated by potentials of different widths $\sigma$, as a function of time. 
It is clearly visible that increasing the width of the spin polarizing potential above the value $\sigma\gtrsim 4\xi$ 
does not lead to the creation of the impurities with larger radii. 
Instead it is observed, that in most cases the size of the impurity remains essentially constant with $r\approx 3.5\xi$. 

\subsection{The impurity stability vs potential strength} \label{appD}
\begin{figure}[t]
\includegraphics[width=\columnwidth, trim=0 140 0 140, clip]{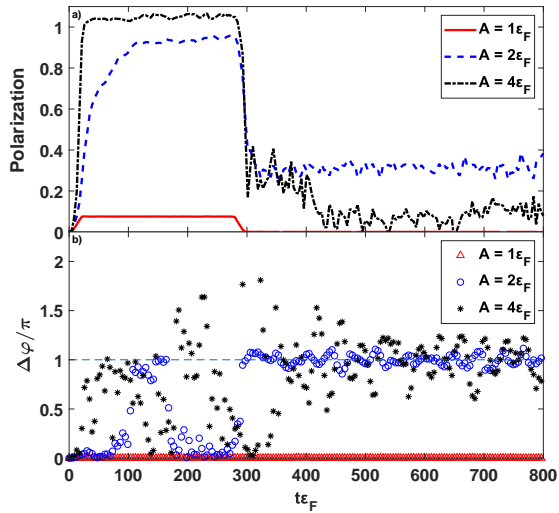}
\caption{ Time evolution of the central local spin polarization (panel a) and the 
pairing phase difference (panel b) for impurities generated by the potentials of different 
strengths. The simulation box is $48^3$ which corresponds to $38\xi$ along in each dimension. The 
width of the gaussian potential is set to $\sigma = 4.71$. 
For three presented scenarios, only the potential of strength $A_0=2\eF$ produces a stable ferron.  \label{SMfig:A_pol_phase}}
\end{figure}

We have analyzed an impact of the potential strength $A_0$ on the process of the impurity formation. 
Time evolution of the spin polarization inside the impurity and the phase difference for three selected strengths are presented in 
Fig.~\ref{SMfig:A_pol_phase}.

If the amplitude is too small $A_0\lesssim\eF$, then the potential is too weak to break efficiently Cooper pairs and to induce locally
a sufficiently large spin polarization. As a consequence the stable impurity is not formed. 
On the other hand, if the amplitude is too large $A_0\gtrsim 4\eF$, the potential induces excitations of the background superfluid
(e.g. phonon excitations), which have sufficient energy to effectively interfere with 
the ferron structure, leading to its decay. Thus there is a particular range of the potential amplitudes for which we observe 
the creation of the long-lived impurity. 

\subsection{Deformed impurities} \label{appE}
\begin{figure}[t]
\includegraphics[width=\columnwidth]{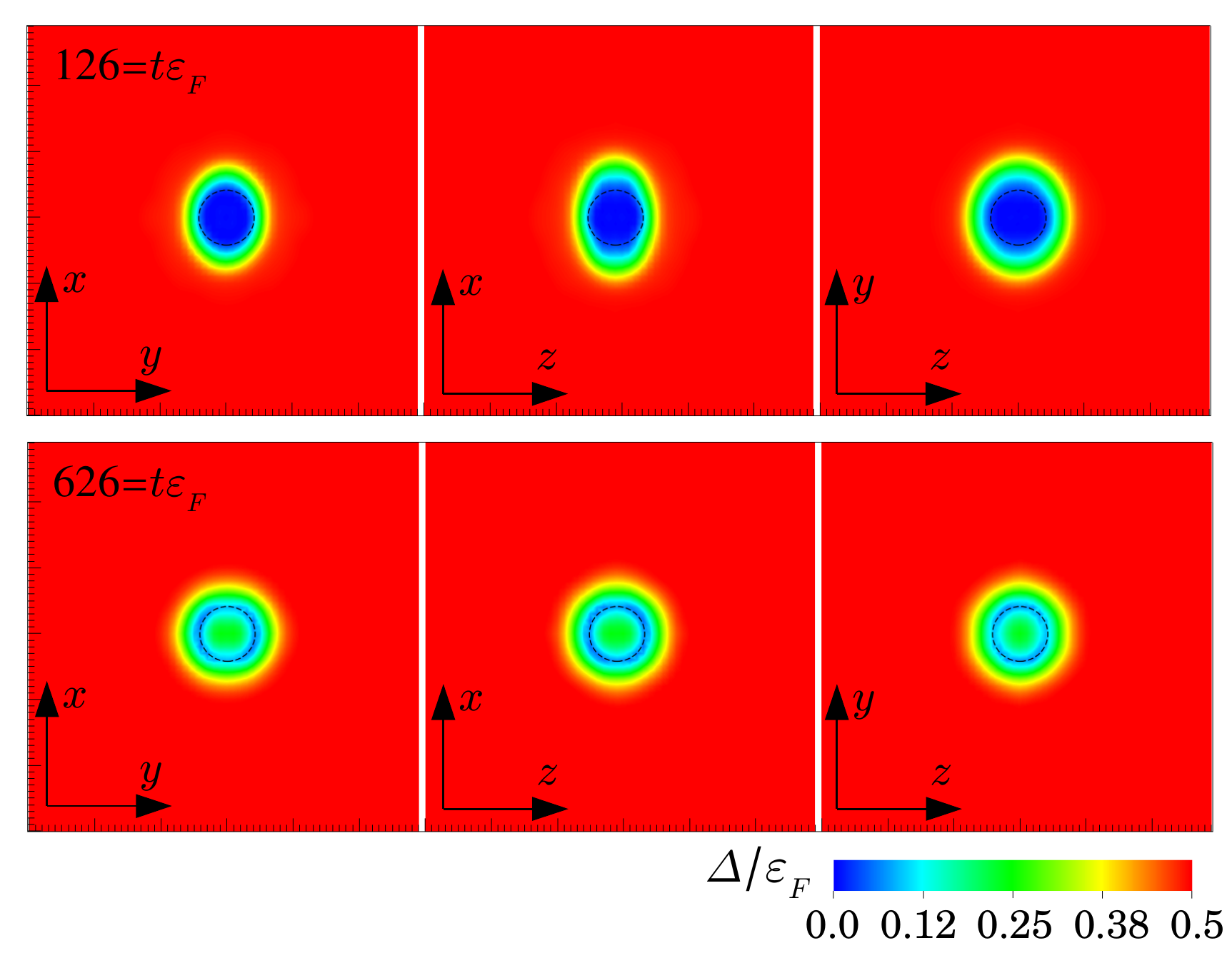}
\caption{Cross sections through the paring field $\Delta(\bm{r})$ along three perpendicular 
planes. The top row subfigures show the pairing field configuration just before the deformed potential ($\sigma=4.7\xi$, $\epsilon_{y}=0.44$ and $\epsilon_{z}=0.64$) is switched off. 
The bottom row subfigures show the pairing field configuration after the potential is switched
off and the system has been evolved for
an additional time interval $500\eF^{-1}$. Black circles (dashed line) are to guide an eye only. The plot indicates 
that spherical shape is a preferable configuration for the ferron.  For full movie see {\it Movie 8}~\cite{Supplemental}. \label{SMfig:deformed}}
\end{figure}

In most of simulations, we were using the spin polarizing potential~(\ref{SMeq:V_p}) with additional, small symmetry breaking parameters 
$|\epsilon_{y/z}|\approx 10^{-6}$. In such a case, almost spherically symmetric impurities were created. 
In order to explore the possibility of creation of a deformed impurity and to investigate its stability, we have executed runs involving
a strongly deformed external potential. Namely, we set potential parameters: $\epsilon_{y}=0.44$ and $\epsilon_{z}=0.64$. 
We have found, that the generated impurity is as stable as the one generated by nearly spherically symmetric potential. 
Moreover, the potential generated initially deformed impurity which subsequently evolved toward the spherical configuration, see Fig.~\ref{SMfig:deformed}. 
It demonstrates that the spherical shape represents the most favorable configuration. 

\subsection{Collision of impurities} \label{appF}
\begin{figure}[t]
\includegraphics[width=\columnwidth]{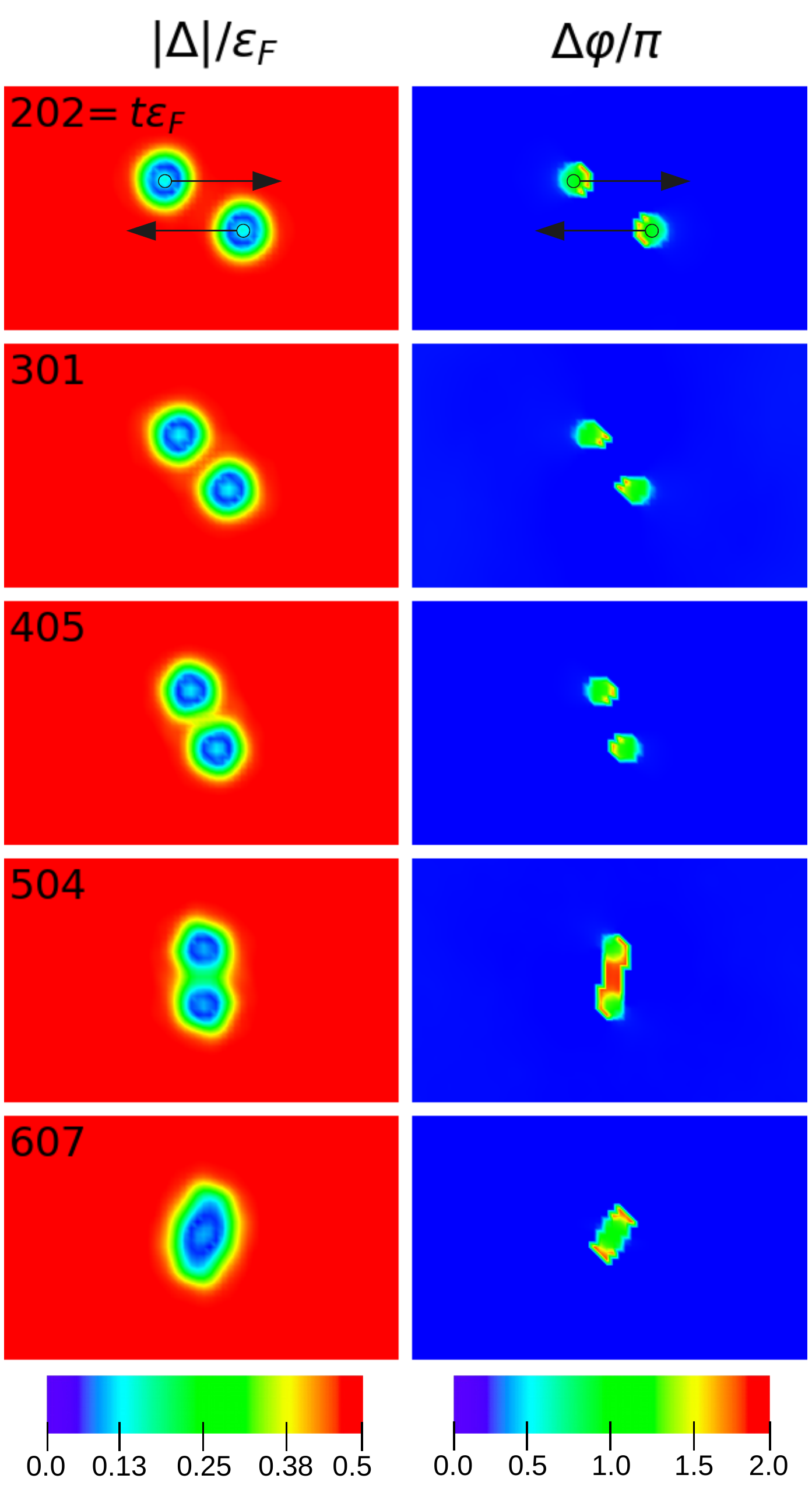}
\caption{ Snapshots from simulation demonstrating the peripheral collision of two ferrons. In the left column 
the absolute value of the paring field is plotted, while in the right column the pairing phase difference 
(with respect to the value at the box edge) is presented. For the full movie see  {\it Movie 9}~\cite{Supplemental}. \label{SMfig:collision_sample}}
\end{figure}
Since ferrons are long-lived and localized excitations, one may consider scenarios which are sensitive to their mutual 
arrangements and investigate their dynamics involving collisions.
There are several issues which may be explored and here we touched upon only the most basic ones. Namely, there is clearly an induced interaction
between ferrons which is mediated by the superfluid background. The natural question is how strong is this interaction and whether it may 
effectively repel or attract impurities. The second question is whether the peculiar structure of ferron is rigid enough to survive collision,
or whether collision leads to its immediate decay.
As examples of such scenarios, we have performed simulations where we collided two ferrons. The ferrons were created by means of two moving potentials 
along $x$-axis. The velocity of potentials was set smaller than the speed of sound. The shape of each of the potential was determined by Eq.~(\ref{SMeq:V_p}).
In Fig.~\ref{SMfig:collision_sample} we have shown few snapshots of a collision with a nonzero impact parameter. 
We have observed a fusion process, which resulted in creation of a new stable ferron. These simulations indicate, that the ferron states represents very stable configurations.

\section{\label{sec4}Experimental realization}

\begin{figure}[h]
   \begin{center}
   \includegraphics[width=\columnwidth, trim=0 0 0 0, clip]{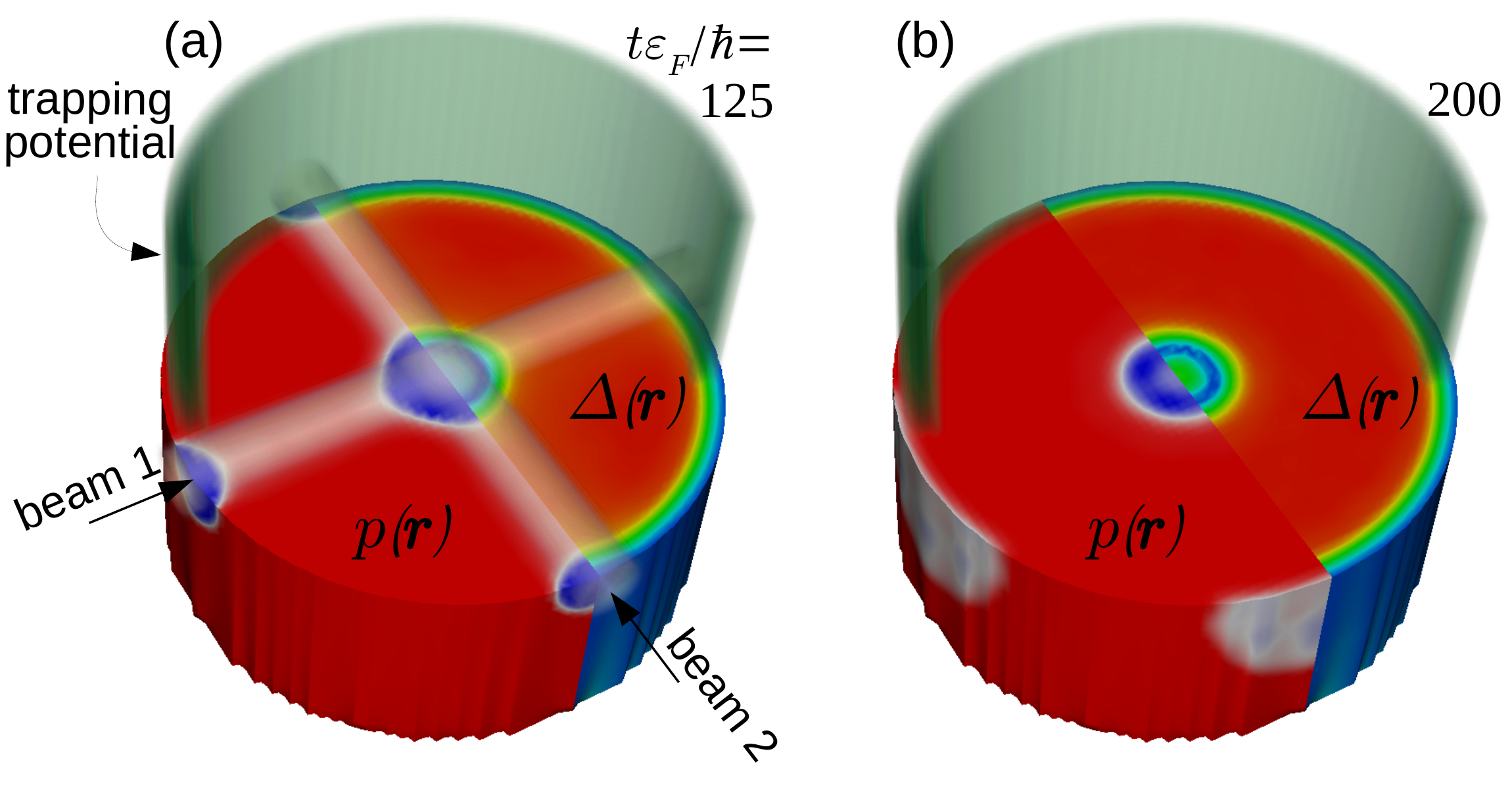}
   \end{center}\vspace{-3mm}
   \caption{Idea of an experimental protocol leading to the ferron creation in a box-like trap. Each panel shows spatial distribution of the local polarization $p(\bm{r})$ and the pairing field $\Delta(\bm{r})$. Color coding is the same as in Fig.~\ref{fig:1}. Two crossing beams, each of amplitude $A_0=1\eF$ and width $\sigma=3.14\xi$ are applied for time $t\eF/\hbar=150$. Only in the crossing region the total strength is sufficient to efficiently polarize the gas, panel~(a). After removal of the beams, the polarized region converts into stable ferron, panel (b). For the full video see the Supplemental \textit{Movie 11}~\cite{Supplemental}. \label{fig:5}}
\end{figure}

Recent developments of experimental techniques allow to implement spin-dependent potentials using  close-to-resonance laser beams~\cite{Lebrat-etal}. Both the size and the strength of the beam can be controlled. In order to induce the ferron~excitation 
it is not sufficient to polarize the system locally -- the spin-polarizing potential has to be strong enough
to allow the pairing field to reappear with the flipped phase.
Clearly, the potential strength must be at least of the order of the paring gap, which is quite large for UFG,  $\Delta/\eF\approx 0.5$. Our calculations indicate that 
the potential of amplitude $A_{0}\approx\eF$ is not sufficient to induce the ferron, and 
one needs $A_{0}\approx 2\eF$ for an efficient droplet creation. Therefore we suggest to use two crossing laser beams, 
each of about $\eF$  amplitude. In the crossing region the amplitude will be enhanced to the required strength, allowing for the creation 
of localized ferron. This procedure has been tested numerically for UFG confined in a box trap~\cite{box-trap}, see Fig.~\ref{fig:5}. 
It has to be emphasized that the associated time scales are also experimentally accessible. In the numerical tests, we were applying the polarizing potential for time intervals of about $150\eF^{-1}$, and we concluded that lifetime of ferrons is at least $1,000\eF^{-1}$. 
Assuming the value of Fermi energy for a typical experimental setup, $\eF\approx h\cdot13$ kHz according to Ref.~\cite{box-trap}, 
the required time for a beam impulse is $2$ ms, which subsequently gives at least $12$ ms for the ferron detection, e.g. by measuring the density difference of two spin states. 

The mechanism responsible for the stability of a ferron is generic and is valid both at the BEC and BCS side. However other features of the system come into play when one considers these limits. Energy needed to create the ferron scales approximately linearly with the pairing gap $E_{imp}\sim|\Delta|$, as the  main energy cost comes from breaking of the Cooper pairs. At the BCS side, the condensation energy scales as $E_{\textrm{cond.}}\sim |\Delta|^2$, and thus 
these two scales will become comparable
in deep BCS regime.
Therefore trying to induce local spin imbalance (the ferron) in the system
one may likely remove the pairing at all, turning the system into the normal phase. 
For more details, see Appendix~\ref{appH}.
In the opposite BEC limit two energy scales exist: the one associated with the binding energy of 
composite bosons, which is steeply rising on the BEC side and the one related to 
the condensation temperature, which is fairly constant.
In order to locally spin-polarize the BEC one needs to break up composite bosons which 
require rather strong laser beams, which would lead to an extensive 
excitation. In our tests (see sec.~\ref{appD}), we find that even at unitarity application of strong potentials (like $A_0\gtrsim 4\eF$) typically induce too much phonons
that effectively interfere with the ferron structure leading to its destruction. 
Thus, we  predict that the unitary regime is the most suitable for the ferron creation. 
There is also another technical problem, namely close-to-resonance laser
beams may heat up the system due to incoherent scattering of photons (effect not included in our simulations) and thus cause the loss of superfluidity before the ferron structure is created. This aspect requires more sophisticated analysis which is beyond scope of this work. 

\section{Impurities with more complex internal structure} \label{appG}

\begin{figure}[t]
\includegraphics[width=\columnwidth]{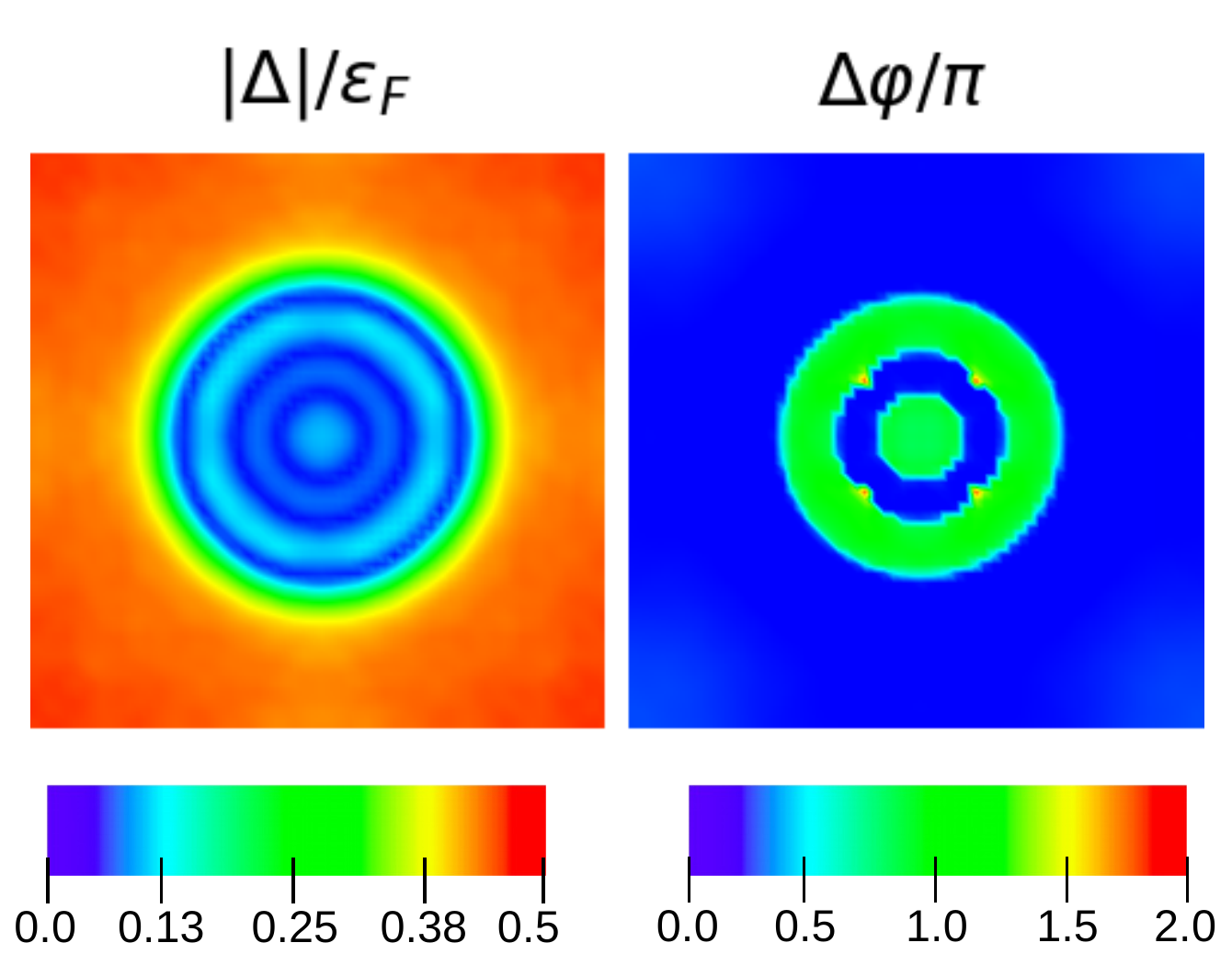}
\caption{Snapshot from simulation demonstrating the internal structure of the large ferron-like excitation taken after the time $\Delta t\approx 220\eF^{-1}$ with respect to the moment when the potential was removed. 
It is clearly seen that the phase changes sign three times as we proceed towards the center of the impurity. For full movie see  {\it Movie 13}~\cite{Supplemental}. \label{SMfig:sigma15_img}}
\end{figure}

Studies of proximity effects for SFS junctions proved that the order parameter may change sign few times, depending on the width 
of the ferromagnet layer~[13,14]. The similar effect is also expected in the case of the ferron-like excitations. 
In order to investigate such a possibility, we have performed exploratory simulations using external potential of large width $\sigma=11.8\xi$ and 
amplitude $A_0=3.5\eF$. 
Indeed, we have observed that after removal of the potential (at time $t_{\textrm{off}}\eF\approx 170$) inside the impurity the order parameter 
changed sign a few time as we move towards the center. Sample configuration is presented in Fig.~\ref{SMfig:sigma15_img}.
The simulation revealed also that the internal pairing structure exhibits various oscillations in time. Further studies are required in order 
to determine whether these oscillations are related to internal dynamics of the impurity or they originate from the superfluid background fluctuations.  

\section{\label{sec5}Conclusions}
We have demonstrated using the state-of-the-art time dependent density functional 
theory that one may create in the bulk of the unitary Fermi
gas, a surprisingly long-lived excitation consisting of  a spin-polarized region
characterized by a peculiar structure of the pairing field, which governs its stability. The mechanism
responsible for its creation is similar to the one responsible for the FFLO
phase, or for the appearance of Josephson-$\pi$ junction in SFS structures.
The experimental conditions for creation of these structures (ferrons) are within a reach  and  may  offer  a  possibility  to  explore
a plethora of new phenomena involving dynamics and interactions
between ferrons and e.g. quantum vortices or domain walls. Note also
that its creation and detection may turn out to be simpler than the detection of FFLO phase 
and thus would provide an indirect strong argument for the existence of this long-sought phase~\cite{VarennaZwierlein}.

\begin{acknowledgments}
We are grateful to Aurel Bulgac and George Bertsch for discussions and critical remarks.
This work was supported by the Polish National Science Center (NCN) under
Contracts No. UMO-2016/23/B/ST2/01789 (PM,BT) and UMO-2017/26/E/ST3/00428 (GW).
We acknowledge PRACE for awarding us access to resource Piz Daint
based in Switzerland at Swiss National Supercomputing Centre (CSCS), decision No. 2017174125. 
We also acknowledge Global Scientific Information and Computing Center, Tokyo Institute of Technology for resources at TSUBAME3.0 (Project ID: hp180066) and Interdisciplinary Centre for Mathematical and Computational Modelling (ICM) of Warsaw University for computing resources at Okeanos (grant No. GA67-14).
The contribution of each of the authors has been significant and
the order of the names is alphabetical.
\end{acknowledgments}

\appendix
\section{Numerical implementation of TDASLDA} \label{appA}
The calculations have been performed
using the numerical code constructed for studies of solitonic cascades in spin-imbalanced unitary Fermi gas (UFG). It was 
described in details in Supplemental Material of Ref.~\cite{SuppressedSolitonicCascade} and therefore here we limit the description to the most important details.
The code solves TDASLDA equations~\cite{LNP__2012} (which have formal structure of time dependent Bogoliubov de-Gennes equations) 
on a 3D spatial lattice without any symmetry restrictions. Periodic boundary conditions are imposed. 
Spatial derivatives appearing in the single-particle Hamiltonian are calculated using spectral methods (via FFTs), 
while for time integration the Adams-Bashforth-Moulton (predictor-corrector) scheme of 5th order was implemented. 
In order to accelerate computations, Graphics Processing Units (GPUs) are utilized. In calculations presented here the number of GPUs 
exceeded $128$ (nvidia P100) and for some cases (for large lattices) it reached $1,600$. 
Calculations were preformed on TSUBAME3.0 (Tokyo Institute of Technology, Japan) and Piz Daint (CSCS, Switzerland) supercomputers.  

We have used lattice sizes ranging from $40^3$ up to $64^3$, with lattice constant $dx=1$. 
Various box volumes allowed us to investigate influence of finite size effects on stability of ferrons. 
In each simulation the number of particles was chosen in a such way to satisfy the condition $\kF=(6 \pi^2 n_{\uparrow})^{1/3}=1$. 
In the case of the smallest box ($40^3$) this requirement corresponds to $1,081$ particles per single spin state. 
The integration time step $\Delta t$ was taken to be $\Delta t=0.005\eF^{-1}$ (units are set by the requirement: $m=\hbar=1$). 
This integration time step enabled to execute numerically stable simulations within time intervals $t_{\textrm{max}}\approx 1,000\eF^{-1}$. 

\section{Protocol for the impurity generation} \label{appB}
\begin{figure}[t]
\includegraphics[width=\columnwidth]{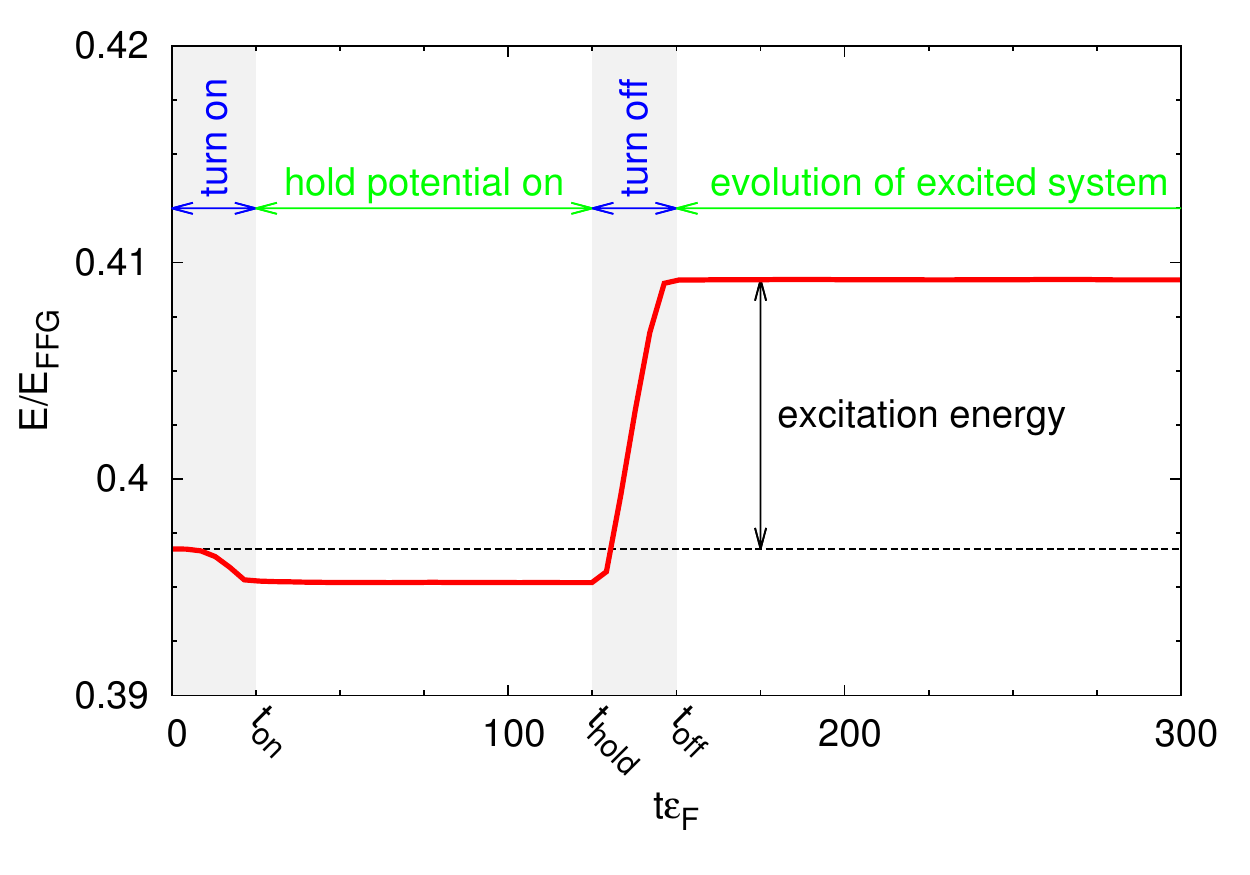}
\caption{ Example of time evolution of the total energy of the system. \label{fig:SM_energy}}
\end{figure}

The initial condition for simulations consists of the uniform solution of unpolarized
unitary Fermi gas, at very low temperature $T/\eF=0.01$. 
The self-consistent solution of ASLDA equations for uniform system provided the correct system properties, namely, $E/E_{\textrm{FFG}}= 0.40(1)$ (Bertsch parameter) and $\Delta/\eF=0.50(1)$, 
where uncertainties arose due to finite size effects, related to various lattice sizes used in the calculations. 
Having the uniform, unpolarized solution of ASLDA, we subsequently applied the spin-selective external potential 
that locally polarizes the system:
\begin{equation}
V_{i}(\bm{r},t)=\lambda_{i} A(t) \exp\left[  -\frac{x^2 + (1-\epsilon_y)y^2 +(1-\epsilon_z)z^2}{2\sigma^2} \right], 
\label{SMeq:V_p}
\end{equation}
where $\lambda_{\uparrow}=+1$ denotes spin-up particles (repulsive potential) and $\lambda_{\downarrow}=-1$ denotes spin-down particles (attractive potential). 
The width of the Gaussian potential $\sigma$ was set to be few times larger than the BCS coherence length $\xi\approx 1.27$. The coefficients $|\epsilon_{y/z}|\ll 1$ were introduced in order to break the spherical symmetry of the potential. 
The amplitude $A(t)$ varied in time according to the prescription:
\begin{equation}
A(t)=\left\lbrace 
\begin{array}{ll}
 A_0\,s(t,t_{\textrm{on}}), & 0\leqslant t<t_{\textrm{on}},\\
 A_0, & t_{\textrm{on}} \leqslant t<t_{\textrm{hold}},\\
  A_0\,[1-s(t-t_{\textrm{hold}},t_{\textrm{off}}-t_{\textrm{hold}})], & t_{\textrm{hold}}\leqslant t<t_{\textrm{off}},\\
  0, & t\geqslant t_{\textrm{off}},
\end{array}
\right. 
\label{SMeq:At}
\end{equation}
where  $s(t, w)$ denotes the function which smoothly varies from 0 to 1 within time interval $[0, w]$:
\begin{equation}
 s(t,w)=\dfrac{1}{2}+
 \dfrac{1}{2}\tanh\left[\tan\left(  \frac{\pi t}{w}-\frac{\pi}{2} \right) \right].
 \label{eq:switch}
\end{equation}
$A_0$ denotes the amplitude of the potential, which we typically set to be about $A_0\approx 2\eF$.
In Fig.~\ref{fig:SM_energy} we present the typical energy evolution of the system being the subject of
the external spin-polarizing potential that is switched on at certain rate and subsequently switched off. 
It is clearly seen that the energy changes only within time intervals when the external potential $V_i$ is turning on and off. 
Otherwise, the total energy of the system is conserved. The energy difference: $\Delta E = E(t_{\textrm{off}})-E(0)$ 
can be attributed to the excitation energy of the system, and it roughly measures the energy contained in the impurity 
(one needs to remember that creation of the impurity in real-time generates also various background oscillations, which
carry part of the energy). 
\begin{figure}[t]
\includegraphics[width=\columnwidth, trim=20 140 20 150, clip]{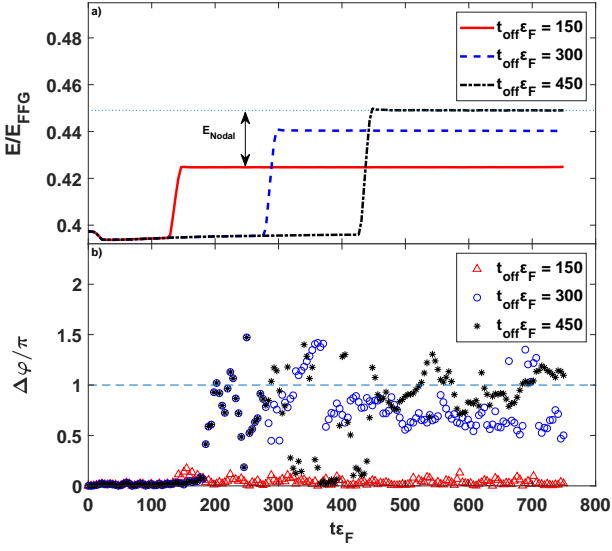}
\caption{Time evolution of the total energy (a) and the pairing phase difference between polarized region and the outside superfluid background~(b). Presented results were obtained for three different time intervals during which the external potential
was turned on. In the first case (red triangles) the potential is applied within the time interval $t_{\textrm{off}}\eF\approx 150$, whereas in the other cases the time interval was enlarged by factor $2$ and $3$, respectively. 
The amplitude of the potential is $A_0=2\eF$ and the width is $\sigma=7.07\xi$. 
Calculations presented in this figure are visualized in {\it Movies 5-7}~\cite{Supplemental}. \label{SMfig:energy_phase}}
\end{figure}

It is important to stress that generation of the stable ferron requires some time. 
If the external potential is applied within too short time interval, then the system will not manage to develop 
a peculiar structure of the pairing field, with the phase shifted by $\pi$ inside the polarized region. 
The situation is demonstrated in Fig.~\ref{SMfig:energy_phase}, where we have compared three types of calculations, differing 
by the time interval during which the potential was kept at its maximum strength.
It is clearly seen, that in the case of the potential which is completely turned off at $t_{\textrm{off}}\approx 150\eF^{-1}$, 
the phase difference between interior of the polarized region and outside superfluid is not created. 
In the second case, when the spin-selective potential was kept on twice longer, 
we have observed that a certain phase difference is developed. It does not reach however the value of $\pi$ and the impurity decays. 
Finally, for the potential duration: $t_{\textrm{off}}\eF\approx 450$ we have
observed that the phase difference $\pi$ is developed (the reason of large fluctuations in this case is discussed in the next section). 
It results in generation of the nodal surface, which strongly suppresses the effects responsible for the decay. 
The energy plot demonstrates that there is indeed a certain energy cost related to the generation of the nodal surface, 
which can be quantified as the total energy difference for the states with and without the flipped phase. 
This contribution to the energy is depicted as $E_{\textrm{nodal}}$ in Fig.~\ref{SMfig:energy_phase}.

\section{Instability of ferrons in 1D} \label{appH}

We argued that the stability of ferrons is attributed to its geometry and dimensionality. Here we present the simple example visualizing the instability of ferron in 1D.
Namely, we have applied TDBdG approach in 1D (all quantities depend on position $x$ and time $t$ that we drop for notation brevity):
\begin{equation}\label{eq:tddft1d}
  i  \frac{\partial}{\partial t}
  \begin{pmatrix}
    u_{n,\lambda}\\ 
    v_{n,-\lambda}
  \end{pmatrix} = 
  \begin{pmatrix}
    h_{\lambda} & \lambda\Delta \\
    \lambda\Delta^*& -h_{-\lambda}^* 
  \end{pmatrix}
  \begin{pmatrix}
    u_{n,\lambda} \\ 
    v_{n,-\lambda}
  \end{pmatrix}.
\end{equation}
where $\lambda=\pm 1$ denotes spin indices, $h_{\lambda}(x,t)=-\frac{1}{2}\frac{d^{2}}{dx^{2}} + g n_{-\lambda}(x,t) + V_{\lambda}(x,t)$ and
$\Delta(x,t) = g\nu(x,t)$ ($n_{\lambda}$ is the density of spin-$\lambda$ particles, 
$\nu$ is the anomalous density). 
The coupling constant $g$ has been adjusted in such a way to fulfill condition: $\Delta/\eF\approx 0.5$, $\eF=\frac{\kF^2}{2}$.
In 1D one may generate the similar form of spin polarized impurity. The {\it Movie 14}~\cite{Supplemental} shows an example of ferron in 1D created by the potential:
\begin{equation}
V_{\lambda}(x,t) = 1.8 f(t) \lambda\eF\exp( -\frac{x^{2}}{2\sigma^2} ),
\end{equation}
where
\begin{eqnarray}
f(t) &=& \sin^{2}(\frac{\pi t}{2T})\theta(T_{1}-t) \nonumber \\
     &+& \theta(t-T_{1})\theta(T_{2}-t) \\
     &+& \cos^{2}(\frac{\pi (t-T_{2})}{2T})\theta(t-T_{2})\theta(T_{1}+T_{2}-t) \nonumber
\end{eqnarray}
describes switching-on and switching-off rates.
For the particular realization of the ferron shown in the {\it Movie 14} we have used parameters: $k_{F}\sigma=4.441$, $T=T_{1}=29.55\eF^{-1}$ and $T_{2}=49.25\eF^{-1}$.
Note that after the initial creation of the ferron with a characteristic structure of the pairing field, 
the polarized regions are repelled from each other 
and the system between them becomes again fully paired. This behavior, which can also be understood by means 
of results presented in Fig.~2 (main article), indicates that ferron is unstable in 1D.

\begin{figure}[h!]
\includegraphics[width=\columnwidth, trim=0 0 0 0, clip]{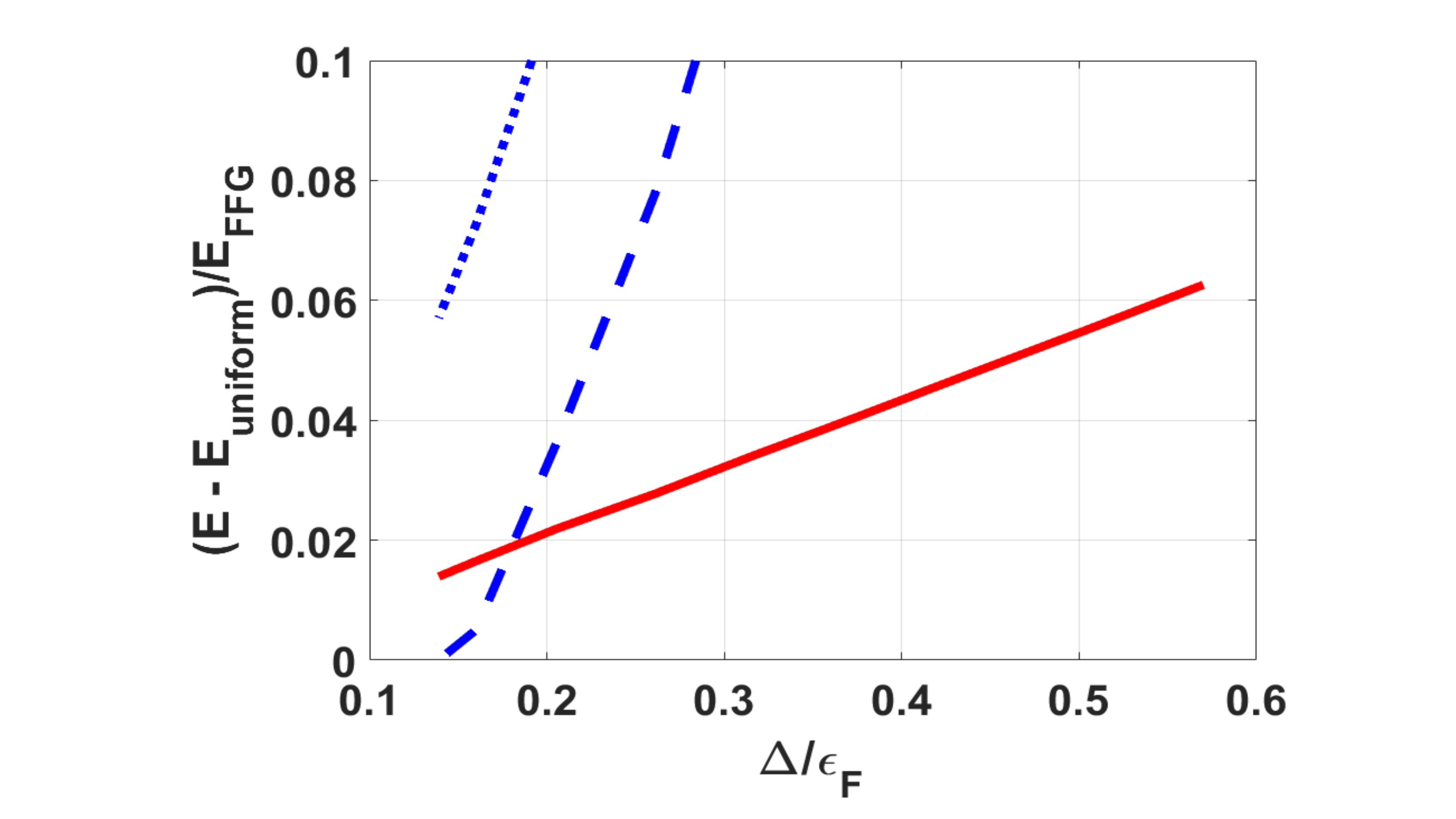}
\caption{The energy of the ferronic structure in 1D (with respect to the uniform
system) vs pairing gap for $N=20$ and $L=200$ (see text for details)
is indicated by the red solid line. The blue dashed line and blue dotted line 
show the pairing energies for
spin imbalanced system: $N+2$ spin-up fermions, $N-2$ spin-down fermions and uniform, unpolarized system,
respectively.
\label{bcs_bec_img}}
\end{figure}

It is also instructive to analyze the structure of the ferron 
and its excitation energy as a function 
of the pairing gap. It sheds light on the properties of the ferron when
moving from the unitary regime towards the BCS limit. We have performed
the analysis in 1D using BdG approach. Namely, we have considered the spin imbalanced
system with $N+2$ spin-up and $N-2$ spin-down fermions
on the lattice of size $L$ with periodic boundary conditions. The spin imbalance generates
two nodes of the pairing field, which we have placed at $x=\pm L/4$. 
It is the only possibility to generate the stationary configuration in 1D.
We have calculated the energy of such configuration with respect to the uniform, unpolarized
system of $N$ spin-up and $N$ spin-down fermions. The results are shown in Fig.~\ref{bcs_bec_img}
as a function of the pairing gap of the uniform system.
It is clearly seen that the energy, which can be thought of as the excitation
energy associated with the creation of the nodal structure of the pairing field 
in spin imbalanced system, changes linearly with the magnitude of the pairing gap.
This is because the 
main energy cost comes from local spin-polarization of the system, ie. Cooper pairs breaking. 
The size of the ferron is much larger in the weak coupling limit due to the increase of the coherence length, which make the 
polarized shell surrounding ferron significantly wider than in the unitary regime.
The dashed lines shown in the figure indicate the behavior of the condensation energy
which scales as $E_{\textrm{cond.}}\sim |\Delta|^2$. In deep BCS regime the energy
required to create the ferron and the condensation energy become comparable and may
even become smaller, since the paring gap decreases exponentially there.
Therefore trying to induce local spin imbalance (the ferron) in the system,
one may likely remove the pairing at all, turning the system into the normal phase.
Consequently in the BCS limit the creation of a ferron may be practically difficult
as it may likely lead to destroying pairing correlations and creating a normal system.

\section{TDASLDA vs BdG description of the ferron} \label{appI}

In the paper we have discussed properties of ferronic excitation obtained within TDASLDA
approach, which was tuned to describe spin imbalanced unitary Fermi gas. The 
analysis indicates that the ferron stability is 
due to the interplay between pairing and spin polarization and is therefore 
generic, not depending on the particular form of the functional. In order to illustrate this feature
we have performed calculations within Bogoliubov-de Gennes (BdG) approach, where only kinetic term 
and pairing term are present (no self-energy contribution). 
The pairing strength has been adjusted to produce pairing gap corresponding to the unitary limit
($|\Delta|/\eF \approx 0.5$).
\begin{figure}[t]
\includegraphics[width=1.0\columnwidth, trim=0 0 0 0, clip]{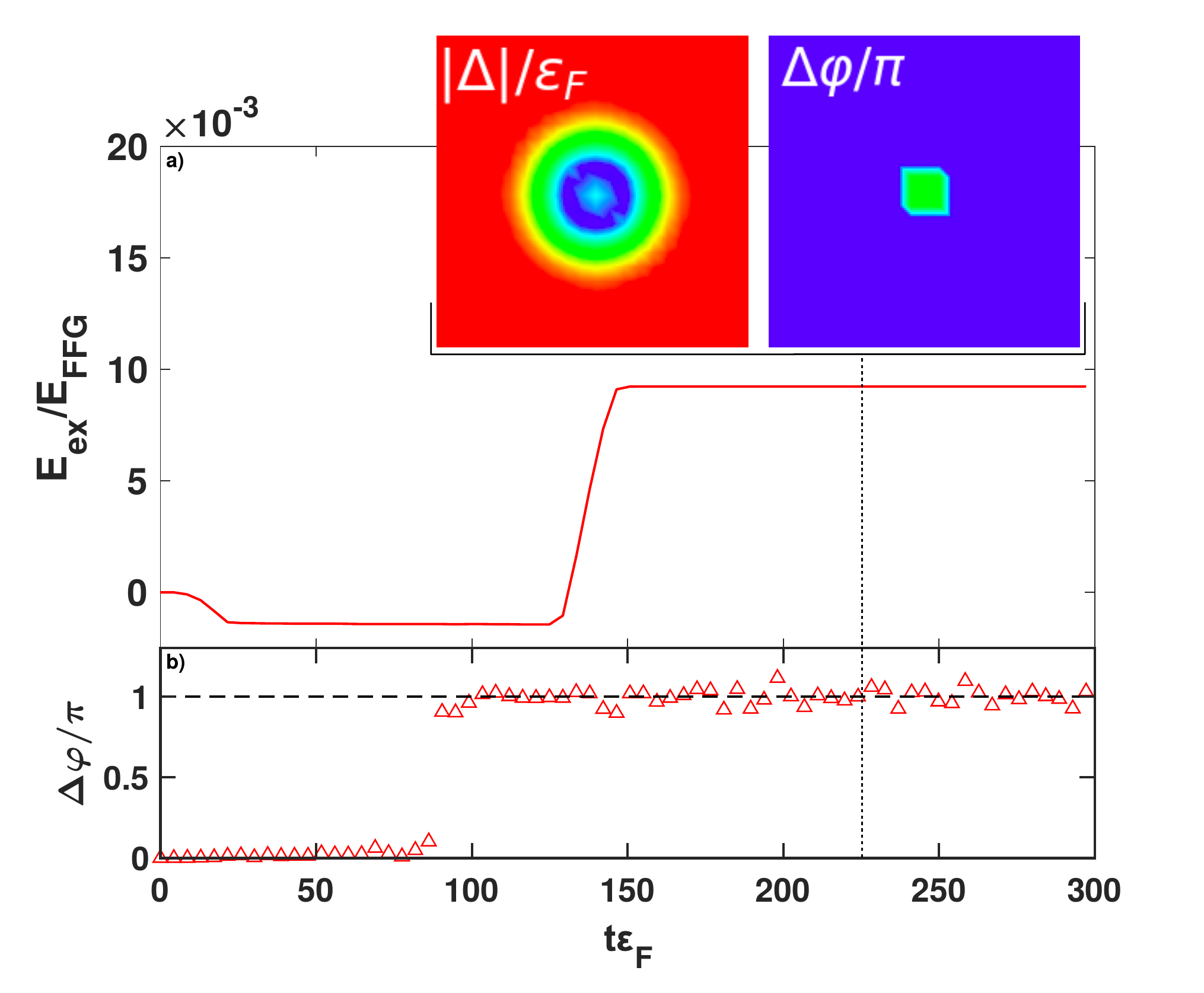}
\caption{Simulation demonstrating the ferron stability
obtained within the BdG approach (see text for details). The total excitation energy $E_{\textrm{ex}}=E(t)-E(0)$ as a function of time is shown together
with the phase difference of the pairing field between the center of the ferron and the surrounding matter. Internal structure of the paring field ($|\Delta|/\eF$) and the phase pattern ($\Delta\varphi/\pi$) is also shown for selected time $t\approx 225\eF^{-1}$. The color coding is the same as in Fig.~9. For full movie see  {\it Movie 15}~\cite{Supplemental}. \label{Bdg:ferron_img}}
\end{figure}
The results are shown in the Fig.~\ref{Bdg:ferron_img}.
Ferron has been created using the same technique as in the case of TDASLDA, ie. by applying
the external potential defined by eq. (\ref{SMeq:V_p}).
The induced local spin polarization, after certain time interval, generates 
pairing phase difference inside the impurity (see the lowest subfigure in Fig.~\ref{Bdg:ferron_img}).  
It is clearly seen that the structure of the ferron is qualitatively the same
as obtained within TDASLDA framework. Moreover the stability of the created object is not affected.
The difference between BdG approach as compared to TDASLDA comes into play only
in the process of the ferron creation. Namely,
the time needed to create the pairing phase flip
inside the impurity, turns out to be different in these approaches.
Summarizing, the existence of a stable ferron is not sensitive to a particular form of the functional.

\newpage
\begin{center}
{\bf Supplemental Material for:}\\
{\bf ``Spin-polarized droplets in the unitary Fermi gas''}\\
\end{center}
{\small The list of supplemental movies, visualizing dynamics of ferrons, is provided.}

\subsection{List of movies}
All movies are also accessible on YouTube. The movies present the distribution of absolute value 
of the paring field $|\Delta(\bm{r})|$, the phase difference of the paring field with respect to the value at the boundary of the box $\Delta\varphi$, 
and the local polarization $p(\bm{r})=\frac{n_{\uparrow}(\bm{r})-n_{\downarrow}(\bm{r})}{n_{\uparrow}(\bm{r})+n_{\downarrow}(\bm{r})}$, in plane crossing the impurity center.

\begin{description}
\item[Movie 1:] Simulation demonstrating the response of the UFG to the time dependent, spherically symmetric, spin polarizing potential defined by Eq.~(B1). Lattice size used corresponds to $40^3$. The amplitude of the potential is $A_0=2\eF$ and the width
$\sigma=2.36\xi$. The potential has been turned on and off within time interval $25\eF^{-1}$ and kept fixed at its maximum strength for $100\eF^{-1}$, exactly as presented in Fig.~12. In this case the width of the potential is too small to produce 
the ferron.\\[2mm]
    File=\verb|movie-01.mp4|,\\
    YouTube=\url{https://youtu.be/5k12xXeYC8E}
    
\item[Movie 2:] The same as {\bf Movie~1} but for the width $\sigma=3.14\xi$. In this case the width exceeds $\sigma\gtrsim 3\xi$ and the stable ferron is created.\\[2mm]
    File=\verb|movie-02.mp4|,\\
    YouTube=\url{https://youtu.be/K6vpatwB7i0}
    
\item[Movie 3:] The same as {\bf Movie~1} but for the width $\sigma=5.5\xi$.\\[2mm]
    File=\verb|movie-03.mp4|,\\
    YouTube=\url{https://youtu.be/1YJhlONwfTQ}
    
\item[Movie 4:] The same as {\bf Movie~1} but for the width $\sigma=6.28\xi$. In this case the potential generates excitations
of the superfluid background which are strong enough to have an impact on the structure of the ferron.    \\[2mm]
    File=\verb|movie-04.mp4|,\\
    YouTube=\url{https://youtu.be/kOt4a29tDn0}
    
\item[Movie 5:] The same as {\bf Movie~1} but for the width $\sigma=7.07\xi$. In this case the time interval during which
the potential is switched on is too short to allow for creation of a ferron.\\[2mm]
    File=\verb|movie-05.mp4|,\\
    YouTube=\url{https://youtu.be/6EsVbUqB4l8}
    
\item[Movie 6:] The same as {\bf Movie~5} but the time interval has been increased to $250\eF^{-1}$, which allows for creation of 
the short-lived ferron-like structure.\\[2mm]
    File=\verb|movie-06.mp4|,\\
    YouTube=\url{https://youtu.be/CQPKSIyBSJ4}
    
\item[Movie 7:] The same as {\bf Movie~5} but the time interval has been increased to $400\eF^{-1}$, which allows for creation of 
stable ferron.\\[2mm]
    File=\verb|movie-07.mp4|,\\
    YouTube=\url{https://youtu.be/pA7P_ps4zE8}
    
\item[Movie 8:] Movie demonstrating evolution of a deformed impurity, see sec. {\it Deformed impurities} for more details. Two selected frames from 
this movie are presented on Fig.~8. The lattice size was set to $60^3$. The amplitude of the potential is $A_0=2\eF$, while 
widths are different for each direction. The potential depends on time in the same way as the one in {\bf Movie~1}. 
The movie demonstrates that spherically symmetric shape of a ferron is energetically preferable.\\[2mm]
    File=\verb|movie-08.mp4|,\\
    YouTube=\url{https://youtu.be/gehgS6YLq_0}
    
\item[Movie 9:] Movie demonstrating peripheral collision of two ferrons, see sec. {\it Collision of impurities} for more details. Panels 
showing the paring field from this movie were used in Fig.~9. The potentials were moving along $x$-axis with velocity $v\approx0.45v_{F}$, where $v_{F}$ is Fermi velocity. The amplitude is $A_0=2\eF$ and width $\sigma=3.14\xi$. The simulation has been
performed on the lattice $64\times 40^2$. \\[2mm]
    File=\verb|movie-09.mp4|,\\
    YouTube=\url{https://youtu.be/C_3Xp-e1m0Q}
    
\item[Movie 10:] The same as {\bf Movie~9}, but for head-on collision. \\[2mm]
    File=\verb|movie-10.mp4|,\\
    YouTube=\url{https://youtu.be/wUY6rJQCHYU}
    
\item[Movie 11:] Movie demonstrating creation of a ferron in a box-like trap. Ferron is created as a result of application of two crossing beams, each of them has amplitude $A_0=1\eF$ and width $\sigma=3.14\xi$. In the crossing region the strength of the potential is sufficiently large to 
induce spin polarization and create a stable ferron. In the calculations the box size $64^3$ was used. Fermi momentum in the trap center 
is $\kF=1$. Note that for regions outside the trap, where the density is $n(\bm{r})\approx 0$ and also $|\Delta(\bm{r})|\approx 0$, 
variations of the phase difference $\Delta\varphi(\bm{r})$ and the local polarization $p(\bm{r})$ are meaningless. \\[2mm]
    File=\verb|movie-11.mp4|,\\
    YouTube=\url{https://youtu.be/cOrsBIZmE2M}

\item[Movie 12:] The same as {\bf Movie~11}, but for the width of the beams $\sigma=4.71\xi$. \\[2mm]
    File=\verb|movie-12.mp4|,\\
    YouTube=\url{https://youtu.be/sz7oabZN99E}
    
\item[Movie 13:] Simulation demonstrating creation of the impurity with more complex internal structure. Lattice size used in the calculation was set 
to $64^3$. The amplitude of the potential is $A_0=3.5\eF$ and width $\sigma=11.78\xi$. The potential has been turned on and off 
within time interval $30\eF^{-1}$ and kept fixed at its maximal strength within $112\eF^{-1}$. Elements of frame at $t\eF=392$ were 
presented in Fig.~11. \\[2mm]
    File=\verb|movie-13.mp4|,\\
    YouTube=\url{https://youtu.be/KfV7f0rljXc}
    
\item[Movie 14:] Creation of a ferron in 1D Fermi gas, described within BdG approximation Eq.~(C1),
using the external potential of the gaussian type with amplitude $A_{0}=1.8\eF$. 
The lattice size is $200$. The potential is turned off after $t=78.8\eF^{-1}$. The expansion of polarized shell is clearly visible 
after switching off the potential.\\[2mm]
    File=\verb|movie-14.mp4|,\\
    YouTube=\url{https://youtu.be/oz13Z00MUZQ}

\item[Movie 15:] Simulation demonstrating creation of the impurity within the BdG approach. Lattice size used in the calculation was set 
to $40^3$. The amplitude of the potential is $A_0=1.5\eF$ and width $\sigma=4.71\xi$. 
The potential has been turned on and off 
within time interval $25\eF^{-1}$ and kept fixed at its maximal strength within $100\eF^{-1}$. Elements of frame at $t\eF=223$ were presented in Fig.~15. \\[2mm]
    File=\verb|movie-15.mp4|,\\
    YouTube=\url{https://youtu.be/NrsKmYm-HRE}

\end{description}

\end{document}